\documentclass[RNAAS]{aastex631}

\begin{document}

\title{Geometric masking in AGN jets and its implications for unification and blazar physics}

\author{Alberto Dom\'inguez}
\affiliation{IPARCOS and Department of EMFTEL, Universidad Complutense de Madrid, E-28040 Madrid, Spain}

\correspondingauthor{A. Dom\'inguez}
\email{alberto.d@ucm.es}

\begin{abstract}
I explore the theoretical implications of the Geometric Masking scenario for AGN Unification and the blazar sequence. In PKS 2155$-$304, GeV spectral hardening appears locked to its $\sim 1.7$\,yr quasi-periodic oscillation trough, suggesting high-flux states are dominated by a soft, geometrically amplified envelope that masks an underlying hard core. While this foundational mechanism builds on observations of coexisting emission regions (Madero \& Domínguez 2026) and phase-locked spectral hardening (Domínguez et al. 2026), here I propose hypothetical extensions that fall outside the scope of those works. I propose that Doppler boosting preferentially enhances this soft component in closely aligned jets. This visibility bias could extend AGN Unification and introduce a geometric modulation layer to the blazar sequence. Under this speculative framework, low-flux states represent windows of geometric transparency, and misaligned systems act as naturally unmasked laboratories. Consequently, the intrinsic duty cycle of extreme acceleration in AGN jets may be higher than inferred from flux-selected observations.
\end{abstract}


\section{1. Introduction}

Blazar gamma-ray emission is largely dominated by stochastic variability \citep[e.g.,][]{Abdo2010, Vaughan2016}. However, a minority of sources show year-scale quasi-periodic oscillations (QPOs), whose origin remains debated between geometric mechanisms and intrinsic plasma instabilities \citep[e.g.,][]{Ackermann2015, Penil2020, Rico2025}. To discriminate between these scenarios, phase-resolved spectral analyses provide an essential diagnostic \citep{Madero2026}.

The Active Galactic Nuclei (AGN) Unification Scheme explains class diversity primarily through orientation \citep[e.g.,][]{Urry1995}, while the blazar sequence relates spectral properties to intrinsic jet power and cooling \citep{Fossati1998,Ghisellini1998}. Though often discussed separately, both geometry and intrinsic physics drive the observed phenomenology.

Recently, evidence for coexisting emission regions was observed in the periodic blazar PG~1553+113 \citep{Madero2026}. Applying a similar analysis to the $\sim 1.7$\,yr QPO in the high-synchrotron peak (HSP) blazar PKS~2155$-$304 revealed that a rare GeV spectral hardening event \citep{Dinesh2025} is phase-locked to the oscillation trough \citep{Dominguez2026}. 

These results suggest support for a Geometric Masking scenario where high-flux states are dominated by a soft, geometrically boosted envelope (the {\it mask}), while plasma-driven hard-spectrum signatures (the {\it core}) become observable when geometric boosting is reduced \citep{Dominguez2026}. Because extending this mechanism to the broader AGN population involves hypotheses far beyond the scope of those initial observational studies, they are presented here to outline directions for future research. In this Note, I examine the hypothetical implications of this framework for AGN unification, the blazar sequence, and acceleration duty cycles.

\section{2. The Physical Mechanism}
The Geometric Masking scenario arises from the interplay between relativistic beaming and the spectral properties of distinct jet components. I assume a two-component structure: (1) a soft-spectrum Mask (sheath/envelope) from cooled electrons, strongly modulated by geometry; and (2) a hard-spectrum Core (spine) from fresh shock acceleration, which is less beamed or intrinsically fainter.

Core visibility is regulated by the Doppler factor $\delta = [\Gamma(1-\beta \cos\theta)]^{-1}$. Flux density scales as $F_{\nu} \propto \delta^{3+\alpha}$ with spectral index $\alpha$. The soft Mask typically has a steeper index ($\alpha_{\rm mask} \sim 1.5$) than the hard Core ($\alpha_{\rm core} \sim 0.5$). For small viewing angles $\theta$ (high $\delta$), the soft Mask amplifies faster ($\propto \delta^{4.5}$) than the hard Core ($\propto \delta^{3.5}$). In aligned high-flux states, differential boosting allows soft emission to dominate, masking the central engine. During geometric minima (QPO troughs or misaligned views), reduced $\delta$ dims the Mask and increases Core contrast, rendering spectral hardening events observable \citep{Dominguez2026}.

In \citet{Dominguez2026}, we suggested that this mechanism might explain the softer-when-brighter behavior in PKS~2155$-$304. We proposed a similar interpretation for PG~1553+113 in \citet{Madero2026}, where flux peaks reflect boosted soft emission rather than fresh particle injection.

\section{3. A Visibility-Regulated Framework for AGN and Blazars}
\subsection{3.1. Implications for the AGN Unification Scheme}
Within the AGN Unification Scheme, orientation distinguishes blazars from misaligned radio galaxies. I extend this: orientation and $\delta$ regulate not just class identity, but which spectral component dominates. Unification may thus function as both a classification and a visibility scheme.

Conventionally, FSRQs exhibit soft gamma-ray spectra due to broad-line region cooling. However, hard intrinsic acceleration signatures may remain concealed beneath boosted soft emission. Integrating over low-flux states (rather than flares) reveals hard spectral components in FSRQs S5~1027+74 \citep{Paliya2025} and 3C~273 \citep{VERITAS2025}. Thus, low-flux states serve as transparent windows where the dimming mask exposes the hard spine, suggesting efficient particle acceleration could be common in FSRQs but observationally obscured.

\subsection{3.2. Implications for the Blazar Sequence}
In the blazar sequence, observed luminosity--spectral trends may partially reflect contrast and selection effects alongside intrinsic cooling. High $\delta$ preferentially boosts the soft component, making aligned sources appear more luminous and spectrally softer, while low-$\delta$ views reveal harder components at lower apparent luminosities. Intrinsic drivers remain fundamental, but geometry introduces a modulation layer shifting sources between masked and unmasked states.

Assuming a soft sheath and hard stochastic spine, as in \citet{Dominguez2026}, the FSRQ/HSP distinction during flares is partly determined by which component geometrically dominates. 

This scenario also provides a natural framework for why FSRQs appear more gamma-ray variable than BL Lacs. In FSRQs, external photon fields amplify intrinsic acceleration via External Compton scattering, producing luminous flares that overpower the mask. In HSPs, flares rely on less efficient Synchrotron Self-Compton processes. These weaker flares remain masked by the boosted envelope, emerging only when the mask dims. Consequently, the intrinsic duty cycle of extreme acceleration may be comparable across classes, despite lower observational detectability in HSPs.

\subsection{3.3. Broader Blazar Phenomenology}
If aligned sources are mask-dominated, misaligned sources (radio galaxies like Cen~A or M87) may be effectively permanently unmasked. With lower $\delta$, soft sheath amplification relative to the hard core drops. Although flux-limited, these sources are contrast-rich; intrinsic hard spectral features appear with a higher duty cycle than in blazars. Observations should prioritize deep spectral integration of misaligned sources to probe acceleration physics.

This helps address a tension with conventional models, where a faster spine ($\Gamma_{\rm spine} > \Gamma_{\rm sheath}$) predicts harder-when-brighter behavior. Following the standard beaming scaling \citep{Urry1995}, differential boosting amplifies the steeper soft component more rapidly, explaining observed softer-when-brighter trends \citep{Dominguez2026, Madero2026}. Rare extreme outbursts can still temporarily outshine the mask.

While low-flux states are typically considered intrinsic quiescence (``engine off''), here they represent geometric transparency windows. Hard components detected during flux minima \citep{Paliya2025, VERITAS2025, Dominguez2026} suggest the intrinsic acceleration duty cycle is high, but observationally regulated by masking.

\section{4. Conclusion}
Rather than a definitive model, the AGN Unification Scheme can be explored as a visibility-regulated framework where orientation and $\delta$ determine both class appearance and jet component contrast. Supported by phase-locked spectral hardening in PKS~2155$-$304 \citep{Dominguez2026} and hard-spectrum detections in low-flux FSRQs \citep{Paliya2025, VERITAS2025}, Geometric Masking indicates high-flux states are often dominated by differentially boosted soft emission.

By treating traditional quiescence in aligned jets as a window of geometric transparency, this framework proposes that misaligned systems serve as unmasked laboratories. The blazar sequence possibly reflects both intrinsic physics and visibility effects; therefore, the intrinsic duty cycle of extreme acceleration may be higher than flux-selected surveys infer.

Future surveys should combine orientation-aware strategies with deep observations of geometric minima to test these hypotheses and probe the exposed central engine.

%\begin{acknowledgments}
%\end{acknowledgments}

\end{document}